\documentclass[preprint]{aastex}

\shortauthors{Nagar et al. 2000}
\shorttitle{VLA survey of 48 LLAGN}
\begin{document}

\title{Radio Sources in Low-Luminosity Active Galactic Nuclei. I. 
VLA Detections of Compact, Flat-Spectrum Cores}

\author{Neil M. Nagar\altaffilmark{1}, Heino Falcke\altaffilmark{2}, 
	Andrew S. Wilson\altaffilmark{1,3}, Luis C. Ho\altaffilmark{4}}
\altaffiltext{1}{Department of Astronomy, University of Maryland, 
		 College Park, MD
                 20742; neil@astro.umd.edu, wilson@astro.umd.edu}
\altaffiltext{2}{Max-Planck-Institut f\"{u}r Radioastronomie, 
	         Auf dem H\"{u}gel 69,
	         53121 Bonn, Germany; hfalcke@mpifr-bonn.mpg.de}
\altaffiltext{3}{Adjunct Astronomer, Space Telescope Science Institute}
\altaffiltext{4}{Observatories of the Carnegie Institution of Washington,
                 813 Santa Barbara St., Pasadena, CA 91101; lho@ociw.edu}
\received{February 22, 2000}
\accepted{April 18, 2000}

\begin{abstract}
We report a high resolution (0\farcs2), 15~GHz survey of a 
sample of 48 low-luminosity active galactic nuclei with the 
Very Large Array{\footnote{The VLA is operated by the National Radio
Astronomy Observatory, a facility of the National Science Foundation 
operated under cooperative agreement by Associated Universities, Inc.}}.
Compact radio emission has been detected 
above a flux density of 1.1~mJy in 57\% (17 of 30) of
low-ionization nuclear emission-line region (LINER) nuclei
and low-luminosity Seyferts. The 2~cm radio power is significantly
correlated with the emission-line ([O~I]~$\lambda$6300) luminosity.
Using radio fluxes at other frequencies from the literature,
we find that at least 15 of the 18 detected radio cores 
have a flat to inverted spectrum 
($\alpha~\geq$ $-$0.3, S$_{\nu}~\propto~\nu^{\alpha}$).
While the present observations are consistent with the radio emission 
originating in star forming regions (the brightness temperatures are 
$\geq$~10$^{2.5-4.5}$~K),
higher resolution radio observations of 10 of the detected 
sources, reported in a companion paper \citep{falet00},
show that the cores are very compact ($\lesssim$~pc), of high brightness
temperature (T$_b~\gtrsim$ 10$^8$~K) and probably synchrotron
self-absorbed, ruling out a starburst origin.
Thus, our results suggest that at least 50\% of low-luminosity Seyferts and
LINERs in the sample are accretion powered, with the radio emission
presumably coming from jets or advection-dominated accretion flows.
We have detected only 1 of 18 ``transition'' (i.e. LINER + H~II) nuclei
observed, indicating their radio cores are significantly weaker than those
of ``pure'' LINERs.

Compact 2~cm radio cores are found in both type 1 (i.e. with broad H$\alpha$)
and type 2 (without broad H$\alpha$) nuclei.
There is weak evidence, limited in 
significance by small numbers, that low-luminosity active galactic nuclei
with compact radio cores exhibit radio ejecta preferentially aligned
along the rotation axis of the galaxy disk. If this result is confirmed by
a larger sample, it would lend support to the idea that the misalignment
of accretion disks with the galaxy stellar disk in more luminous Seyfert
galaxies is a result of radiation-pressure induced warping of their 
accretion disks.
\end{abstract}


\section{Introduction}

There is increasing evidence that a large fraction of 
the nuclei of nearby galaxies show many similarities with 
powerful active galactic nuclei (AGN); these objects are termed 
low-luminosity active galactic nuclei (LLAGN; active galaxies 
with nuclear L$_{H\alpha}~\leq$ 10$^{40}$ erg s$^{-1}$;
Ho, Filippenko \& Sargent 1997a, hereafter H97a).
These similarities include broad
H$\alpha$ lines (Ho et al. 1997b), broader H$\alpha$
lines in polarized emission than in total emission
\citep{baret99}, nuclear UV sources \citep{maoet95,baret98}
and water vapor megamasers \citep{braet97}.
However, the emission-line spectra of LLAGNs
(i.e. low-luminosity Seyferts, LINERs, and ``transition'' nuclei 
[nuclei with spectra intermediate between those of
LINERs and H~II regions]), can also be modeled in terms of photoionization 
by hot, young stars \citep{termel85,filter92,shi92}, by collisional 
ionization in shocks  \citep{kosost76,foset78,hec80,teret92,dopsut95}, 
or by aging starbursts \citep{aloet99}.
Thus, it is not clear that accretion onto a black hole powers all LLAGNs.

How does one distinguish accretion-powered LLAGNs from
LLAGNs powered by hot stars or supernova shocks? 
Broad H$\alpha$ lines and bright 
unresolved optical or UV sources are ambiguous 
indicators because they can all be produced in starburst
models (see Terlevich et al. 1992), and a search for
broader polarized H$\alpha$ emission is currently feasible in 
only a few of the brightest LLAGNs \citep{baret99}.
Further, all these indicators 
are highly dependent on viewing geometry and obscuration and on the 
signal to noise of the observations, a problem exacerbated
by the low optical and UV luminosities of LLAGNs. Their observation 
demands both high signal-to-noise spectra and high spatial resolution to
separate weak nuclear emission lines from the starlight of the host galaxy.

One well-known property of some powerful AGNs is a compact,
flat-spectrum nuclear radio source, usually interpreted as
the synchrotron self-absorbed base of the jet which fuels 
larger-scale radio emission. Astrophysical jets are known to
be produced in systems undergoing accretion onto a compact
object \citep[e.g.][]{pri93,bla93}, so such compact
radio sources in galactic nuclei may reasonably be considered
a signature of an AGN. Much theoretical work 
\citep[e.g.][]{beget84,lovrom96,falbie99} has been devoted
to this disk-jet relationship in the case of galactic nuclei
and it has been suggested that scaled-down versions of AGN jets
can produce flat-spectrum radio cores in LLAGNs
\citep{fal96,falbie96}.
Nuclear radio emission with an inverted spectrum
is also expected from an advection-dominated accretion flow 
\citep[ADAF; e.g.][]{naret98}, a possible form of accretion 
onto a black hole at low accretion rates \citep{reeet82}, or from
bremsstrahlung, cyclotron and synchrotron emission from
plasma accreting quasi-spherically onto a black hole
\citep{mel94}. 
From the observational
perspective, \citet{hec80} showed that LINER nuclei tend to
be associated with a compact radio source, and compact,
flat-spectrum radio cores are known to be present in many
`normal' E/S0 galaxies \citep{sadet89, wrohee91, sleet94}. 
Flat-spectrum radio cores are, however, uncommon in normal spirals
or Seyfert galaxies \citep{ulvwil89,vilet90,sadet95}. 
Flat-spectrum radio sources can also result through thermal emission
from optically-thin ionized gas or through free-free
absorption of non-thermal radio emission, a process which probably
occurs in compact nuclear starbursts \citep{conet91}. 
The brightness temperature, T$_b$, in such starbursts is limited to
log~[T$_b~$(K)]~$\lesssim$ 5 \citep{conet91}. Thus it is 
necessary to show that T$_b$ exceeds this limit before 
accretion onto a black hole can be claimed as the power source.

In this paper, we present a high-frequency (15~GHz or 2~cm), high-resolution 
($\sim$~0{\farcs}15) survey of LLAGNs with the 
Very Large Array \citep[VLA,][]{thoet80}.
This resolution is high enough to isolate nuclear emission
from that of the host galaxy, and the radiation is unaffected by
the obscuration present at UV or optical wavelengths,
and less affected by free-free absorption than observations at
longer cm wavelengths. Further, large samples can be quickly studied,
as deep radio maps are achievable in as little as 
15--20 min per object.
Higher resolution follow-up observations are then required in order
to eliminate the possibility that the radio cores are thermal in
origin, as discussed above.
We have therefore embarked on a program to observe a large number of 
LLAGNs at high resolution with the VLA and the Very Long Baseline 
Array \cite[VLBA,][]{napet94} in order to identify 
accretion-powered nuclei, to test the predictions of ADAF and 
jet models, and to characterize the presence and structure of radio jets
on sub-pc to hundred-pc scales.
In this paper, we report on the results of the first stage of this
program --- VLA observations of 48 LLAGNs which
have been extensively observed at other wavebands.
Preliminary results of this project have been published in
\citet{falet97,falet98,falet99} and \citet{naget99b}.

\section{Sample, Observations and Data Reduction}

A total of 48 galaxies were selected from the magnitude-limited 
Palomar spectroscopic survey of 486 nearby bright galaxies 
(Ho, Filippenko, \& Sargent 1995).
All 48 galaxies were originally selected, using preliminary spectral
classification, to have nuclei with ``pure'' LINER or ``transition object''
(composite LINER~+~H~II) characteristics. 
The spectral type of 8 of the nuclei was somewhat uncertain, and 
these galaxies were later reclassified as Seyferts (H97a). Thus, 
our sample of 48 LLAGNs consists of 22 ``pure'' LINERs, 18 ``transition
objects,'' and 8 low-luminosity Seyferts. The 48 objects were not 
selected by well-defined criteria; they were individually chosen
for study at multiple wavelengths because they were particularly
``interesting'' LLAGNs. Our sample is the subject of a 
rigorous observational campaign involving radio observations at
2~cm (reported here), 3.6~cm and 6~cm \citep{hoet00}, 
{\it Hubble Space Telescope (HST)} 
optical \citep{penet98} and UV imaging \citep{maoet95,baret98} as well as
spectroscopy \citep{maoet98}, and X-ray imaging by the {\it Chandra}
observatory.

All 48 LLAGNs in the sample, plus two additional
objects -- NGC~4449 (whose nuclear spectrum is that of an 
H~II region) and NGC~5756 (a southern spiral not in the Palomar
spectroscopic survey) -- were observed by the VLA
over three runs on 1996 October 7, 11 and 18,
while the array configuration was being changed from `D' to 
`A' (see Thompson et al. 1980 for a description of the 
VLA configurations).
The weather was fair for all runs with clear skies and temperatures
between 0 and 23{\arcdeg}C. 
Two 8 min observations of each galaxy
were sandwiched between three 2 min observations of a nearby
phase calibrator, so that the total time on source was typically
16 min.
For the October 7 run, there were 9 antennas at A-configuration
pads plus 18 antennas at D-configuration pads.  
Five D-configuration antennas were taken out of the array at various 
times during the run
to be moved to their A-configuration pads; they did not start observing
again until after our run was completed.
A sixth D-configuration antenna was removed from the array for painting.
Of the nine antennas at A-configuration pads, the data from the three
most recently moved were unusable, since the antenna
positions were poorly determined.
For the October 11 run, 25 of the 27 antennas were on their A-configuration
pads, but the data from the four most recently moved antennas were
unusable as their positions were poorly determined.
For the October 18 run, all except one of the antennas were at their 
A-configuration pads, and all had reasonably well-determined positions.

The data were calibrated using the AIPS software, following the standard
procedures outlined in the AIPS cookbook. Since positions for
recently-moved antennas were not well determined at the time of observations, 
antenna positions listed in the data files were updated with the 
corrections to the positions found by VLA operators over the few weeks after 
the observations.
We found that using all position corrections
up to those determined on November 23, 1996, led to the best calibrator 
phase solutions for most of the recently-moved antennas. 
Other bad data were removed 
before the phase solutions, derived from the phase calibrators, were applied
to the target galaxies.
On the October 7 and October 18 runs, the source 1328+307 (3C~286) was 
observed twice 
and used to set the flux-density scale to that of \citet{baaet77}.
On the shorter October 11 run, an observation of 0404+768 was used to set
the flux-density scale (we used a 2~cm flux-density of 1.46~Jy for
this source).
The phase calibrator flux-densities were bootstrapped to that of
0404+768, and the bootstrapped values were found to be in good agreement
with those listed in the VLA flux-density database 
(part of the ``VLA calibrator manual''; available on-line
at http://www.nrao.edu).

The maps were made with AIPS task IMAGR. For the October 7 run, we 
imaged each object twice, once using only the antennas on D-configuration pads,
which resulted in $\sim$5{\arcsec}-resolution maps, and once using only 
those baselines which included at least one antenna on an 
A-configuration pad, which resulted in 
$\sim$0{\farcs}15 resolution maps. For the other two runs, a single 
map was made for each source, using all antennas. 
For sources stronger than about 3~mJy, we were able to
iteratively self-calibrate the data so as to increase the signal-to-noise
ratio in the final map. Phase-only self-calibration was found to give 
good results, with little further improvement from a combined amplitude 
\& phase self-calibration.
The typical r.m.s. noise in the maps prior to self-calibration was 0.1~mJy.
Many of the maps of the undetected nuclei show 4 to 5~$\sigma$ 
peaks at various locations in the map.
We therefore conservatively use 10 times the r.m.s. noise 
in the map as the upper limit for the flux density of the undetected sources,
so the sensitivity of our survey is about 1.1~mJy.

On the October 11 run, NGC~266, NGC~404, and NGC~7217
were also observed at 0.7~cm and 3.6~cm, with the VLA split into
two subarrays which observed at different frequencies
(not all antennas had 0.7~cm receivers).
These data were reduced as described above.
An observation of 0404+768 was used to set the flux-density
scale at 0.7~cm. We set the 0.7~cm flux density of 0404+768 to 578.2~mJy 
for the AC IF's and 575.3 for the BD IF's, after finding, from the VLA 
flux-density database, that these were the flux densities on 
1997 February 24, the date closest to our run that the source was observed
(the flux density stayed between 500 and 600~mJy for the following 
$\sim$300 days). The flux density of the phase calibrators,
0055+300 and 2201+315, were bootstrapped to that of 0404+768,
and the results were consistent with the values listed in
the flux-density database.
At 3.6~cm, the observation of 3C~48 could not be used to set the 
flux-density scale, as the four innermost antennas (and almost every
alternate antenna on each arm) were observing at 0.7~cm, so
that none of the 3.6~cm interferometer baselines was small enough
for an accurate flux-density calibration solution (the VLA calibrator
manual specifies that A-configuration observations of this calibrator should
be restricted to 0--40 k$\lambda$).
We therefore set the flux density of 2201+315
(the phase calibrator for NGC~7217)
to 2.51~Jy in both AC and BD IF's at 3.6~cm, as this was its 
measured value on 1997 January 17, the date closest to our run
(after this date, the flux density of this source rose steadily to 3.5~Jy
over the next $\sim$300 days and then stayed roughly constant).
The flux density of 0055+300
(the phase calibrator for NGC~266 and NGC~404)
was then bootstrapped from that of 2201+315. The 
result, S(0055+300)~= 0.82 Jy, is consistent with the values listed
in the flux-density database for this source.
Nevertheless, the 3.6~cm and 0.7~cm flux densities of the three galaxies 
are subject to additional uncertainty.

All data were calibrated and imaged independently by two of us (NMN and HF), 
yielding consistent results (all corrections for antenna positions were 
also independently done). We are therefore confident of the results, 
despite the uncertainties caused by the uncertain antenna positions.

\section{Results}

\subsection{Detection Rate}

The results of the VLA observations are listed in Table 1, 
with columns explained in the footnotes. 
The radio positions for the 22 detected objects
are limited by the positional accuracy of the phase calibrators,
which is typically 2--10~milli-arcsec (mas),
and by the accuracy of the Gaussian fit to the source brightness distribution, 
which depends on the signal to noise ratio of the source detection.
The overall accuracy should typically be better than $\sim$50~mas for all 
target sources detected in A-configuration, and better than 500~mas for target
sources detected in only D-configuration.
We have compared the radio positions derived here with optical positions 
from Cotton, Condon, \& Arbizzani (1999), which were measured from the digital
sky survey with typical 1$\sigma$ accuracy 1{\farcs}5--2{\farcs}5 in
each of right ascension and declination.
For all but two of the 2~cm detected nuclei, the radio and optical positions
agree to within 1{\arcsec}--3{\farcs}5 (see Table 1). In
NGC~185 and NGC~4550, the offsets between radio
and optical positions are significant, and as discussed in Appendix A,
it is possible that the radio sources we detected in these two galaxies
are not related to the galaxy nucleus.

Assuming that the radio sources towards NGC~185 and NGC~4550 are in fact
associated with their nuclei, the detection rate of 2~cm radio cores 
in the A-configuration maps of our LLAGN 
sample is 37\% (18 of 48), at a 10$\sigma$ detection limit of 1.1~mJy. 
The rate is more striking if we consider the detection rates 
for low-luminosity Seyferts (75\%; 6 of 8) and for LINERs (50\%; 11 of 22).
On the other hand, only one of the 18 observed `transition'
nuclei was detected. The morphological type of the galaxy with
the detected transition nucleus, NGC~5866, is one of the earliest of 
the transition objects observed.
An additional four objects were detected in D- but not A-configuration maps;
we do not consider these four as nuclei with compact 2~cm radio cores.
In comparison, \citet{wrohee91} detected 30\% (64 of 210)
of all E/S0 galaxies with D $<$ 40 Mpc and m$_B$ $<$ 14 mag
in a 6~cm VLA survey with resolution 5{\arcsec} and
10$\sigma$ detection limit 1~mJy;
the lower resolution and lower frequency make their
survey more prone to detecting emission from sources unrelated
to the central engine.

The radio luminosities of most of the detected 2~cm cores are between
10$^{19.3}$ and 10$^{22}$ W Hz$^{-1}$ (Fig.~1), similar to the
luminosities seen in `normal' E/S0 galaxies 
\citep{sadet89}. It is notable, however, that a significant fraction of
the detected 2~cm compact cores are in spiral galaxies (Fig.~1).

The 0.7~cm maps of NGC~404, NGC~266, and NGC~7217 were, as expected, noisy,
and all three galaxies were undetected at a 10$\sigma$ limit of 10~mJy.
At 3.6~cm, NGC~266 and NGC~7217 were detected as unresolved
sources with flux densities 3.1~mJy and 1.2~mJy, respectively, 
while NGC~404 was not detected at a 10$\sigma$ upper limit of 0.9~mJy.

\subsection{Compact, Flat-Spectrum Cores}

None of the objects show reliable extended structure in either
A- or D-configuration  maps.  This is not surprising as the high resolution 
may resolve out most extended emission, which is expected to be weak
at the high frequency observed. For a few sources, a Gaussian fit 
(with peak flux-density, major and minor axes as free parameters)
to the nuclear radio source brightness distribution 
results in a Gaussian size slightly larger than the beam size so that
the peak flux-density is slightly smaller than the total
flux-density (see Table 1);
however, the deconvolved source sizes are much smaller than half the beam
size, so we consider these sources as unresolved.
Of the 8 objects which are detected in both our A- and D-configuration maps,
NGC~2655 is the only one with significantly  more flux density at 5{\arcsec} 
resolution than at 0{\farcs}15 resolution.
Thus, most of the detected 2~cm nuclear radio sources are compact at
the 0{\farcs}15 resolution (typically 15--25~pc) of our survey.
The implied brightness temperatures for the 2~cm compact
nuclei are T$_b$~$\geq$ 10$^{3.0-4.0}$~K except for 
NGC~4278 and NGC~6500 which have T$_b$~$\geq$ 10$^{4.5}$~K, 
and NGC~185, NGC~4548, NGC~4550, NGC~4636 and NGC~5033 which 
have T$_b$~$\geq$ 10$^{2.5-3.0}$~K.

We have supplemented our 2~cm observations with radio data at other
wavelengths from the literature, such as the 6~cm radio survey of
\citet{wrohee91} and the 20~cm VLA FIRST survey catalog 
\citep[hereafter FIRST]{whiet97}, in order to derive 
non-simultaneous spectral indices for the 2~cm detected nuclei;
details of the spectral index determination are listed in Appendix A
for each galaxy detected at 2~cm.
While the actual value of the spectral index is uncertain,
given the resolution mismatch and the non-simultaneity of the 
observations, we can determine whether the core is 
flat spectrum ($\alpha \geq~-$0.3; S$_\nu \propto \nu^{\alpha}$),
or steep spectrum ($\alpha <~-$0.3), as noted in column 13 of Table 1. 
Since compact flat-spectrum radio sources are often variable, 
the use of non-simultaneous data at two frequencies
can cause some intrinsically flat-spectrum radio 
sources to be misclassified as steep-spectrum sources.
However, since extended steep-spectrum radio sources are not
variable, the use of non-simultaneous data at two frequencies
should rarely cause intrinsically steep-spectrum radio sources to be 
misclassified as flat-spectrum sources. Also, since the resolution
is better at the higher frequency, resolution effects will tend to
steepen the measured spectrum if extended emission is present. 
Thus we are confident of the reality of the flat spectra obtained.

Fifteen of the 18 objects detected in our A-configuration
2~cm maps show 
a flat-spectrum radio core, and one has an undetermined radio spectrum.
The remaining two objects detected in the A-configuration 2~cm maps,
NGC~2655 and NGC~4636, 
show evidence for the presence of a steep-spectrum radio core,
although NGC~4636 has ``jet''-like steep-spectrum  nuclear radio extensions 
\citep{stawar86} which may dominate over any flat-spectrum nuclear component
within the beam of the radio maps.
In the case of NGC~2655, the inferred steep-spectrum of the nucleus
is consistent with the presence of extended 2~cm radio emission.

The radio morphology of the LLAGNs detected at 2~cm is different 
from what is seen in ``classical'' Seyferts \citep{ulvwil89,naget99a},
where the nuclear radio emission is usually dominated by steep-spectrum 
``jet'' or double radio components on tens to hundreds of pc scales,  
and any pc-scale radio core may potentially be hidden by
free-free absorption at cm wavelengths \citep[e.g.,][]{ulvet81,galet99}.
The high incidence of compact, flat-spectrum radio cores in the spiral 
galaxies in our sample is unusual \citep[e.g.,][]{vilet90,sadet95}.
In fact, the high incidence of flat-spectrum radio cores in 
low-luminosity Seyferts is also unusual, given that flat-spectrum compact
radio cores are found in only about 10\% of ``classical'' Seyferts
\citep{debwil78,ulvwil89,sadet95}.

\subsection{The Relationship Between Radio and Emission-line
Luminosities}

Correlations between the emission-line and radio luminosities of active
galaxies are well known, and are found in both Seyfert \citep{debwil78}
and radio galaxies \citep{bauhec89}. 
The [O~III]~$\lambda$5007 line is usually used in assessing these correlations,
but this line is weak in LINERs and we use instead [O~I]~$\lambda$6300, which
also has the advantage of being uncontaminated by emission from H~II regions.
In our LLAGN sample, the 2~cm radio power appears closely correlated with the
nuclear [O~I]~$\lambda$6300 luminosity (Fig.~2a and 2b). 
Testing the statistical significance of this apparent correlation is
not straightforward as there exist both upper limits (line not detected)
and highly uncertain values (non-photometric data) for the 
[O~I]~$\lambda$6300 luminosities, 
along with upper limits to the 2~cm radio powers.
Eight of the LLAGNs in the sample have non-photometric 
[O~I]~$\lambda$6300 flux densities. We treated these eight flux densities as
photometric measurements when using the bivariate tests
from the ASURV statistical software package 
\citep[][hereafter ASURV]{lavet92}.
This approximation should not bias the results significantly as the 
uncertainties in the flux density are expected to be much less than an 
order of magnitude.
When this is done, tests on the whole sample indicate a correlation between 
the radio and [O~I]~$\lambda$6300 luminosities at the 97\% significance level.
Deleting NGC~185 (the point at the bottom left corner of Fig.~2a), for
which the association of the radio source with the nucleus is 
uncertain (Section 3.1), lowers the significance level by only 1\%.

\subsection{The 2~cm Non-detections}

Do the LLAGNs undetected at 2~cm truly lack radio cores, or do a significant
fraction of them have radio cores which are weaker than those in the
detected LLAGNs because of, for example, a lower bulge 
luminosity \citep{sadet89} or a later morphological type? 
The distributions of host galaxy bulge luminosity and distance 
(as listed in H97a) are more or less similar for the detected and 
undetected LINERs.
The 2~cm compact core LINERs have a median host galaxy morphological type
(T$_{median}$= $-$1.5) which is earlier than, and a median
[O~I]~$\lambda$6300 luminosity (Log~L$_{[O~I]}$(median)~= 31.5 Watts)
which is higher than, that for LINERs without compact
2~cm cores (T$_{median}$= 2.0, Log~L$_{[O~I]}$(median)~= 31.1 Watts),
though the difference is not statistically significant. 
Given the correlation between emission-line and radio luminosity,
it is possible that deeper radio imaging will find
compact radio cores in the 2~cm undetected LINERs. 
The host galaxies of the two undetected Seyferts 
have bulge luminosities in the B band among the lowest of the eight
Seyferts in our sample. 
There is no clear difference between the 
morphological types of the detected and undetected Seyferts.

Seventeen of the non-detected nuclei are ``transition objects.''
Statistical tests from the ASURV package
indicate that the distributions of the 2~cm radio power of LINERs and
transition objects in the sample are different at a confidence level 
of 99.3--99.5\%. While the numerical value of the significance level is 
questionable given that there is only one detection among the 
transition objects, arbitrarily changing the four transition objects with the
lowest radio flux-density upper limits into radio detections still results in a 
difference between transition object and LINER radio powers at the
92--98\% significance level. 
While the transition nuclei have somewhat lower median bulge luminosities 
and generally later morphological types compared to the detected LINERs,
the statistical significance of these differences is not high. 
Not unexpectedly, the transition objects in our sample do appear 
to have weaker [O~I]~$\lambda$6300 luminosities as compared to 
the LINERs and Seyferts (Fig.~2).
If the non-photometric [O~I]~$\lambda$6300 flux measurements are considered
to be accurate, statistical tests from the ASURV package indicate that in our 
sample the [O~I]~$\lambda$6300 luminosities of the transition 
objects are lower than that of LINERs and Seyferts at the $\sim$99.8\% 
significance level. It remains to be seen whether this luminosity difference
is large enough to explain the striking difference in the detection rates.
Surprisingly, the one transition nucleus detected at
2~cm, NGC~5866 (the rightmost circle in Fig.~2b), has one of the 
lowest [O~I]~$\lambda$6300 luminosities among the transition nuclei.

\subsection{Other AGN indicators}

We find compact 2~cm radio cores in both type 1 and type 2 nuclei, 
that is, those with and without visible broad H$\alpha$ emission.
Furthermore, some type 1 nuclei in our sample (two Seyferts and
four LINERs) do not have a compact radio core at our detection limits.
Among LINERs, 64\% (7 of 11) of type 1 nuclei host compact radio cores, as 
compared to only 36\% (4 of 11) of type 2 nuclei. 

Additional evidence supports the accretion-powered nature of the nuclei
with compact, flat-spectrum 2~cm cores.
NGC~4579 is one of the very few LLAGNs with a detected Fe~K$\alpha$ line
(Terashima et al. 1998, 2000); such lines are believed to originate 
through fluorescence in an accretion disk. The optical--UV spectrum of 
NGC~4579 has been studied in detail by \citet{baret96}, 
who show that the broad-line region in this object is detectable in a number
of permitted transitions. The broad-band spectral energy distribution of 
NGC 4579 \citep{ho99} has been interpreted in the context of an ADAF 
model \citep{quaet99}.
NGC~4278 and NGC~6500 are both known to harbor pc-scale radio cores studied 
with Very Long Baseline Interferometry (VLBI) techniques 
\citep{jon84,jonet81,falet00}, and NGC~4203 \citep{shiet00}, NGC~4278 
and NGC~4636 \citep{maget98} show kinematic 
evidence for massive dark objects at their centers.
\citet{maoet98} find that
the nuclei of NGC~404, NGC~4569, and NGC~5055, all undetected at 2~cm,
contain massive stars which probably generate sufficient ionizing photons 
to power the optical emission lines, so it is not necessary to invoke an
accretion-powered source. On the other hand, they find that the nuclei of 
M~81, NGC~4594, and NGC~6500 (and perhaps NGC~4579) 
exhibit a severe ``deficit'' in ionizing photons 
(i.e. apparently insufficient photons shortward of the Lyman limit, as 
inferred through extrapolation of the observed UV spectrum, to power the
optical and UV emission lines), but their X-ray/UV ratios are two orders 
of magnitude higher than those of the three stellar-powered LINERs. 
\citet{maoet98} concluded that, in these four objects a component that emits 
primarily in the extreme-UV may be the main photoionizing
source. We observed and strongly detected two of these nuclei
(NGC~4579 and NGC~6500) at 2~cm, and the other two are known to have
pc-scale radio cores \citep{baret82,graet81}. 
Finally, the lack of clear evidence for X-ray emission from an AGN in 
NGC~404, NGC~4111, NGC~4192, 
and NGC~4569 \citep{teret99}, galaxies undetected at 2~cm, is consistent
with an absence of an accretion-powered source.

\subsection{Minor-axis Jets or Outflows}

A significant number of the LLAGNs with 2~cm compact radio
cores in our sample show extended 20~cm radio emission
in 5{\arcsec} resolution FIRST maps,
and/or 1{\farcs}5--4{\arcsec} maps from other sources
(see Appendix A). 
We therefore compared the P.A. of this radio extension,
when available, with that of the host galaxy major axis for all LLAGNs 
in our sample which host 2~cm compact radio cores. The relevant P.A.'s
are listed in Table 2, and a histogram of the difference in the
two P.A.'s, $\delta$, is shown in Fig.~3. The single transition 
object (shown in grey) shows extended emission along 
the host galaxy major axis, so it is likely that the radio emission is from
the galaxy disk. On the other hand, LINERs and low-luminosity Seyferts
with compact 2~cm radio cores appear to favor extended radio emission 
along the galaxy disk minor axis, although the small number of objects 
prohibits any meaningful statistical tests.
The extent of this minor-axis radio emission is 
typically only 0{\farcs}5 to 3{\arcsec}
(50--350~pc for most galaxies in the sample),
and could trace the inner parts of wide-angle, minor-axis outflows, as
seen in classical Seyferts \citep{colet96}, or 
low-luminosity collimated ``jets'' from the central engine. 

\section{Discussion}

We have detected unresolved 2~cm nuclear radio emission in 75\% of 
low-luminosity Seyferts and 50\% of LINERs. At least 83\% of these nuclear
radio sources have flat radio spectra. At this point it is not clear whether
the 2~cm non-detected LINERs and low-luminosity Seyferts really do not have
compact radio cores, or whether their compact cores simply have lower radio
luminosities or are hidden by, for example, free-free absorption.
In powerful AGNs, compact flat-spectrum nuclear radio cores are widely accepted
as the signature of synchrotron emission from the base of a synchrotron
self-absorbed jet (see the review by Zensus 1997). 
The compact, flat-spectrum radio cores we detect in LLAGNs could be 
scaled-down versions of powerful AGNs \cite[e.g.][]{falet99}, 
or could trace emission from an ADAF \citep{naret98}, or material undergoing 
quasi-spherical accretion \citep{mel92}. Alternatively, the radio cores
could be produced by thermal bremsstrahlung from 
ionized gas or by synchrotron radiation which has suffered free-free
absorption. Such emission could originate from an active 
nucleus or nuclear star forming regions.
The lower limits to the brightness temperatures, T$_b^{2cm}$, of the 2~cm 
compact core objects in our sample are 10$^{2.5-4.5}$~K, so the 
VLA observations alone cannot rule out the notion that the radio emission 
originates in star formation regions. 
However, most objects have brightness temperatures too high to be 
optically-thin thermal emission from a gas at 10$^4$~K. The expected value of 
T$_b$ is $\sim$ 10$^4$~K ($\nu/\nu_{\tau})^{-2.1}$, where $\nu$ is the
observing frequency and $\nu_{\tau}$ is the frequency at which the
source becomes optically thick. 
Many of our objects have a flat spectrum between 2~cm and 20~cm suggesting
$\nu_{\tau}$~$<$ 1.4~GHz (20~cm). If this is the case, then thermal emission 
must have T$^{2cm}_b < 10^4 (20/2)^{-2.1}$ K
i.e. $<$~90~K, lower than all of our measurements.
Further, higher resolution observations
of 10 of the detected objects in this sample reveal brightness
temperatures $\gtrsim$~10$^8$~K, strongly suggesting that synchrotron
self-absorption is responsible for the flat spectra and arguing against
a thermal origin \citep{falet00}.
Thus, it is very likely that at least 50\% of the LINERs and 
low-luminosity Seyferts in our sample contain accretion-powered nuclei.
Several independent factors, described in Section 3.5, support our
proposition that the presence of a compact, flat-spectrum (as determined 
from VLA observations) radio core is related to the presence of accretion 
onto a massive black hole in LLAGNs.

The detection rate of radio cores in transition objects in our
sample is dramatically low (1 of 18).
The significant difference between transition and ``pure'' LINER detection
rates, combined with the lack of any significant difference between their
host galaxy properties, suggests one of the following. 
1) The central engines of transition
objects are intrinsically different from those of ``pure'' LINERs, or
2) Transition objects are truly composite nuclei with
a LINER and an H~II component \citep[e.g.][]{hoet93}, in which case both 
the LINER component and its radio emission are much weaker in transition
objects than in ``pure'' LINERs.
Using the luminosity in the [O~I]~$\lambda$6300
line as an indicator of the luminosity of the LINER component, this
second premise is supported by the results that the
[O~I]~$\lambda$6300 and 2~cm luminosities are correlated at the
$\sim$97\% significance level, and that the [O~I]~$\lambda$6300
luminosity of the transition objects is lower than
that of the LINERs and Seyferts in the sample at the $\sim$99.8\%
significance level (Fig.~2).

Finally, if the extended 20~cm radio emission, which appears
preferentially aligned
with the host galaxy minor axis in a subset of 2~cm compact core LLAGNs
(Fig.~3), does trace jets from the central engine, there are interesting
implications for the orientation of the central engines of AGN. 
Similar analyses of ``classical'' Seyferts 
\citep{ulvwil84,schet97,claet98,nagwil99} 
have shown that the distribution of 
P.A.$_{radio}$ -- P.A.$_{galaxy~major~axis}$
is more or less uniform between 0{\arcdeg} and 90{\arcdeg}. 
One explanation for this uniform distribution is that radiation from the
central engine illuminates the inner accretion disk inducing a 
``radiative instability'' \citep{pri97} which causes large warps in the 
inner disk. In this case, even if the outer accretion disk shares the
same axis of rotation as the host galaxy disk, the jet
(presumably launched along the axis of the inner accretion disk)
may then be at a random angle with respect to the axis of the galaxy disk.
If the radiation instability is indeed the correct explanation,
then in LLAGNs, where the radiation field of
the central engine is much weaker, the disk is expected to be less
prone to the radiation instability (see e.g. equation 3.2 and Section 3.2
of Pringle 1997), so that the jet axis is more likely to be along the
host galaxy rotation axis. This is precisely what we find. 
Alternately, the jet power in LLAGNs may be low enough that buoyancy 
dominates jet motions on the scale (typically 50--350~pc) of the extended
emission, redirecting the jet along 
the minor axis.  Our further surveys for 2~cm radio cores 
in LLAGNs and our completed observations of 17 nearby bright LLAGNs at 
0{\farcs}3 resolution with the Multi-Element Radio Linked Interferometer
Network (MERLIN) at 20~cm will potentially reveal more on this subject. 

\section{Conclusions}

Our detection of compact, flat-spectrum radio cores in
about 50\% of low-luminosity Seyferts and LINERs in a sample of 48 
low-luminosity AGNs, when combined with VLBA observations \citep{falet00},
suggests that at least half of all low-luminosity Seyferts and LINERs are 
accretion powered. Given the sensitivity limit of our survey, the true 
incidence of radio cores is likely to be higher.
The 2~cm radio power is significantly correlated with the 
[O~I]~$\lambda$6300 luminosity for LLAGNs in our sample. 
The lower detection rate (at $\geq$92\% significance) of compact radio
cores in ``transition'' nuclei suggests that either these nuclei are 
intrinsically different from ``pure'' LINERs, or their ``pure''
LINER component is of lower luminosity, with correspondingly lower
luminosity radio cores.
The latter interpretation is favored by the correlation between the radio
and [O~I]~$\lambda$6300 luminosities coupled with the lower [O~I]~$\lambda$6300
luminosities of the ``transition'' nuclei as compared to the LINERs.

The presence or absence of a detected broad H$\alpha$ line is 
not a good indicator of the presence or absence of a compact, 
flat-spectrum 2~cm radio core.
A significant number of the LINERs and low-luminosity Seyferts 
which contain 2~cm compact radio cores show evidence, at 
low resolution (1{\arcsec}--5{\arcsec}) and frequency (1.4~GHz or 20~cm),
for extended radio emission along the galaxy disk minor axis. 
This radio emission may trace a 
wide-angle outflow or a weak, highly-collimated jet along the disk 
rotation axis. If the latter is true, it lends support
to the idea that it is the ``radiative instability'' which
causes warps in the nuclear accretion disks of more luminous
Seyfert galaxies.

Finally, we note that the data presented here are the initial
results of a larger program to study a well-defined sample
of LLAGNs at high resolution with the VLA and VLBA; results of
these will appear in future papers in this series.

\acknowledgements
This research has been supported by NSF grant AST~9527289 and by NASA
grant NAG~81027. 
The work of HF is funded by DFG grant Fa 358/1-1 \& 2.
The work of L.~C.~H. is partly funded by NASA grant NAG 5-3556, and by NASA 
grants GO-06837.01-95A and AR-07527.02-96A from the Space Telescope Science 
Institute (operated by AURA, Inc., under NASA contract NAS5-26555).

\appendix
\section{Notes on Individual Objects}
Specific notes on all galaxies detected at 2~cm and on some
undetected galaxies are listed here.

\paragraph{NGC~185}
The 2~cm source we detect is offset 14{\farcs}5 west and
4{\arcsec} north of the optical position determined by
\citet{cotet99} from digital sky survey plates. The 
1{$\sigma$} error of the optical position determination is estimated
to be 2{\farcs}7 in each of right ascension and declination. 
The galaxy is diffuse, $\sim$ 12{\arcmin} x 10{\arcmin} in extent, and has 
a prominent dust lane to the north-west which may bias the optical position 
determination. Overlaying the optical and radio images shows that the 2~cm 
radio source lies at the south-west tip of the dust lane, and the offset from
the optical center does appear to be more than can be explained
by dust lane obscuration. Further, the radio source has a very low
luminosity (10$^{16.7}$ Watts Hz$^{-1}$) which can be explained by 
a source in the galaxy stellar disk.
\citet{hum80} lists a 5$\sigma$ upper limit of 
10~mJy in a 23{\arcsec} resolution map at 20~cm,
and we find a 2~cm flux density of 0.8~mJy in our 0{\farcs}15
map. It is therefore possible that the source is flat spectrum,
but a measurement at a resolution closer to that of the 2~cm 
map is required for a definitive result, and for
now we consider this source to have an undetermined
spectral index. 

\paragraph{NGC~266}
Our simultaneous 3.6~cm and 2~cm observations on 1996 October 11,
with resolutions 0{\farcs}29 x 0{\farcs}24, and
0{\farcs}15 x 0{\farcs}13,
detected an unresolved nuclear source with flux densities 
3.1~mJy and 4.1~mJy, respectively. 
The nucleus therefore has an inverted radio spectrum, 
$\alpha^{3.6}_{2}~\geq$ 0.5. 
This source was, however, not detected at a 10$\sigma$ upper
limit of 10~mJy in our simultaneous 0.7~cm observation.
NGC~266 has been identified
with source NVSS~J004947+321637 by the NASA Extragalactic Database
\citep[NED;][]{helet91} i.e. it has been detected
in the 20~cm NRAO VLA sky survey (NVSS), with a peak flux-density of 
8.9~mJy/beam at a resolution of 45{\arcsec}, though the  20~cm
position is offset 44{\arcsec} from our 2~cm position.
In any case, 8.9~mJy may be taken as an upper limit to the 
20~cm flux density.

\paragraph{NGC~404}
Our simultaneous 3.6~cm, 2~cm, and 0.7~cm observations on 1996 October 11 
did not detect the nucleus at 10$\sigma$ upper limits of 
0.9~mJy, 1.3~mJy, and 10~mJy, respectively.

\paragraph{NGC~2655}
Comprehensive radio continuum and optical emission-line imaging, 
as well as optical spectra, are presented by \citet{keehum88}.
Their VLA 6~cm map shows a roughly E-W symmetric 
extension at 0{\farcs}5 resolution 
(with a peak flux-density of 36~mJy/beam from a component
with half-power size 0{\farcs}4 x 0{\farcs}5). 
Combining this peak flux-density with our 2~cm peak flux-densities of 
6~mJy/beam and 13.4~mJy/beam at resolutions of 0{\farcs}15 and 5{\arcsec},
respectively, results in a non-simultaneous spectral index, $\alpha^{6}_{2}$~= 
$-$0.7 to $-$1.6, for the radio emission within 0{\farcs}5 of the nucleus.
\citet{keehum88} find 
S$_{\nu} \propto \nu^{-0.67}$ for the nucleus,
from 4{\arcsec} resolution non-simultaneous
VLA maps at 6 and 20~cm.
They also find a secondary 6~cm source in P.A. 117{\arcdeg}, with 
a diffuse bridge of radio emission joining it to the nuclear component. 
The morphology, kinematics, and line ratios of the surrounding
emission-line gas, suggest that this secondary 6~cm source and 
bridge are related to nuclear ejecta \citep{keehum88}. 
Clearly, most of the radio emission in this object is extended,  
and if the emission from the unresolved radio nucleus is related to an
accreting black hole, then its steep radio spectrum suggests that it is more
likely that the emission is from synchrotron-emitting jets, 
rather than from a compact radio core.
The host galaxy P.A. is not defined for this galaxy in the 
Uppsala General Catalogue of Galaxies \citep[][hereafter UGC]{nil73}
or the Third Reference Catalogue of Bright Galaxies
\citep[][hereafter RC3]{vauet91} as the galaxy is nearly round.

\paragraph{NGC~2681}
FIRST lists a peak 20~cm flux density of 3.8~mJy/beam at 5{\arcsec}
resolution, with extended emission in P.A. 35{\arcdeg}.
We did not detect this object at 2~cm.
The UGC and RC3 do not list a P.A. for this galaxy as it
is almost round.

\paragraph{NGC~2787}
\citet{hecet80} find 20~cm and 6~cm peak flux-densities
of $<$9~mJy/beam and 9~mJy/beam, at resolutions of
4{\arcsec} and 1{\farcs}7, respectively,
so the nucleus has a flat or inverted spectrum between
20~cm and 6~cm, $\alpha^{20}_{6} \geq$ 0.
Combining the 6~cm peak flux-density with our 2~cm peak flux-densities of 
7~mJy/beam and 8.9~mJy/beam at resolutions of 0{\farcs}15 and 5{\arcsec}, 
respectively, results in a non-simultaneous 
spectral index, $\alpha^{6}_{2}$ between $-$0.2 and 0, for 
all emission within the central 1{\farcs}7.

\paragraph{NGC~3147}
\citet{vilet90} find S$_{20cm}^{peak}$ = 10.6~mJy/beam
and S$_{6cm}^{peak}$ = 8.55~mJy/beam at resolutions of $\sim$1{\farcs}2
and $\sim$1{\farcs}5, respectively, with both
maps showing only a compact core. 
Consistent with this, \citet{lauet97} find a nuclear peak
flux-density of 9~mJy/beam at 6~cm at a resolution of 0{\farcs}4.
Combining this 6~cm flux density with our 2~cm peak flux-density 
(8~mJy/beam), the non-simultaneous spectral index of the emission
in the central 0{\farcs}4 is flat, $\alpha^{6}_{2}~\geq$~0. 

\paragraph{NGC~3169}
\citet{humet87} find a core flux-density of 8.2~mJy 
in the central 2{\arcsec} of their
1{\farcs}2 resolution, 20~cm map, and also extended 
emission in P.A. 120{\arcdeg}, more or less along the minor
axis of the host galaxy. Combining their 20~cm peak flux-density
with our measured 2~cm peak flux-densities of 6.8~mJy/beam and 10~mJy/beam
at resolutions of 0{\farcs}15 and 5{\arcsec}, respectively,
results in a non-simultaneous spectral index, $\alpha^{20}_{2}$,
between $-$0.1 and 0.1, for the radio emission within the central 2{\arcsec}.

\paragraph{NGC~4111}
FIRST lists this object's peak 20~cm flux density  as
5.8~mJy/beam at 5{\arcsec} resolution. The
emission is extended in P.A. 70{\arcdeg}, along the minor
axis of the host galaxy.
We did not detect this object at 2~cm.

\paragraph{NGC~4143} -
\citet{wrohee91} find a 6~cm flux density of 6.7~mJy at 5{\arcsec}
resolution.  The similar resolution (5{\arcsec}) FIRST map shows a peak 
flux-density of 5~mJy/beam at 20~cm, with extended emission in P.A.
10{\arcdeg}, which is not along the minor axis of the host galaxy.
Combining the FIRST peak flux-density with our 2~cm peak flux-densities
of 3.3~mJy/beam and 5.3~mJy/beam at resolutions of 0{\farcs}15 and 5{\arcsec}, 
respectively, results in a non-simultaneous spectral index, $\alpha^{20}_{2}$
between $-$0.2 and 0, for all emission within the central 5{\arcsec}.
\citet{hoet97b} note that broad H$\beta$ may also be present in this object.

\paragraph{NGC~4203}
\citet{wro91} derived a 6~cm flux density of 12.5~mJy at
$\sim$5{\arcsec} resolution,
and FIRST lists a peak 20~cm flux density of 6.9~mJy/beam 
(at a resolution of 5{\arcsec}),
implying that the core has a highly inverted spectrum
between 20~cm and 6~cm, $\alpha^{20}_{6}~\sim$
0.5. Our D-configuration 2~cm flux density of 12.3~mJy, suggests a 
non-simultaneous spectral index $\alpha^{6}_{2}~\sim$ 0.
The extended emission in the FIRST map is in P.A. 
90{\arcdeg}, more or less
along the minor axis of the host galaxy.

\paragraph{NGC~4216}
\citet{humet87} observed this object with 1{\farcs}3 
resolution at  20~cm
and did not detect any emission in the central 2{\arcsec}
of the galaxy, at a 3$\sigma$ upper limit of 0.5~mJy.

\paragraph{NGC~4278}
\citet{jon84} have observed this object
using VLBI and find the core flux density is 180~mJy and 190~mJy at
18~cm and 6~cm, on size scales $<$5~mas and $<$1~mas,
respectively. \citet{wilet98} find a 3.6~cm peak flux-density of 153~mJy/beam
at 200 mas resolution, and we find a 2~cm peak flux-density of 88.3~mJy/beam at 
150~mas resolution, so the core may be flat between 18 and 6~cm and
then steeper down to 2~cm. Alternatively, the lower flux density at 2~cm 
may be due to source variability.
The 20~cm emission detected in
FIRST is extended in P.A. 166{\arcdeg}, but the UGC and RC3 do
not list a host galaxy P.A. as the galaxy is almost round.

\paragraph{NGC~4419}
\citet{humet87} find a core flux-density of 32~mJy in the
central 2{\arcsec} of their
1{\farcs}3 resolution 20~cm map, and extended
emission in P.A. 77{\arcdeg}. Combining this flux density with
our 2~cm D-configuration peak flux-density of 5.4~mJy/beam at 5{\arcsec} 
resolution, results in a non-simultaneous spectral index, 
$\alpha^{20}_{2}~\leq -$0.8. This object was not detected 
in our 0{\farcs}15 resolution, 2~cm A-configuration map.

\paragraph{NGC~4449}
FIRST lists a 20~cm peak flux-density of 1.7~mJy/beam at 5{\arcsec}
resolution, with extended emission in P.A. 65{\arcdeg},
more or less along the host galaxy major axis.
We did not detect this object at 2~cm.

\paragraph{NGC~4527}
\citet{vilet90} find S$_{20cm}^{peak}$ = 4.3~mJy/beam
and S$_{6cm}^{peak}$ = 1.3~mJy/beam at resolutions of $\sim$1{\farcs}3
and $\sim$1{\farcs}4, respectively (a spectral 
index, $\alpha^{20}_{6}$, of about $-$1.1).
The 20~cm and 6~cm maps are
both highly extended (P.A. 65{\arcdeg}) along the major axis of the
galaxy, with an edge brightened morphology, 
suggestive of an annulus \citep{vilet90}.
We did not detect this galaxy at 2~cm.

\paragraph{NGC~4548}
\citet{humet87} find a 3$\sigma$ upper limit of
2.5~mJy at 20~cm, over the central 2{\arcsec}
of their 1{\farcs}3 resolution VLA map.
Since we detect a flux density of 1.2~mJy at 2~cm, the 
non-simultaneous spectral index of the core is
$\alpha^{20}_{2} \geq$ $-$0.3.

\paragraph{NGC~4550}
The 2~cm source we detect is offset 9{\farcs}3 west and
5{\farcs}5 north of the optical position determined by
\citet{cotet99} from digital sky survey plates. The 
1{$\sigma$} error of the optical position determination is estimated
to be 2{\farcs}3 in each of the R.A. and Dec. 
The galaxy does not show any obvious sign of a peculiar morphology,
and the east-west galaxy extent is $\sim$1{\arcmin}, so the 
radio-optical offset  does appear significant.
A $\sim$3{\arcsec} resolution 11~cm map \citep{haysra80}
shows an unresolved radio core at a position between our radio
detection and the optical nucleus.
\citet{wrohee91} find the 6~cm emission over the
central 5{\arcsec} is $<$~1~mJy. We measure a 
2~cm peak flux-density of 0.7~mJy/beam, so the source is probably flat 
spectrum, $\alpha^{6}_{2} \geq$ $-$0.3.

\paragraph{NGC~4565}
\citet{humet87} find a core 20~cm flux density of 1.2~mJy in the
central 2{\arcsec} of their 1{\farcs}3 resolution VLA map. 
FIRST lists a peak flux-density of 1.5~mJy/beam at 20~cm in a 5{\arcsec} 
resolution map, with extended emission in
P.A. 36{\arcdeg}, along the minor axis of the host galaxy.
Combined with our 2~cm peak flux-density of 3.7~mJy/beam at 0{\farcs}15 
resolution, the non-simultaneous spectral index is
$\alpha^{20}_{2}~\geq$ 0.5.
This galaxy has the lowest broad H$\alpha$ luminosity found 
in any known active nucleus (Ho et al. 1997b).

\paragraph{NGC~4579}
\citet{humet87} find a core flux-density of 25.1~mJy in the
central 2{\arcsec} of their
1{\farcs}3 resolution, 20~cm VLA map, with extended
emission in P.A. $\sim$135{\arcdeg}.
The radio core was found to have an inverted spectrum between 13~cm and
3.6~cm with the Parkes-Tidbinbilla 275 km interferometer
(S$_{\nu}~\propto$ $\nu^{0.2}$; Sadler et al. 1995).
Van der Hulst, Crane \& Keel (1981) 
found the 6~cm emission to be extended
in P.A. 134{\arcdeg}, with extent 0{\farcs}4$\pm$0{\farcs}2 and
flux density 39.9~mJy.

\paragraph{NGC~4636}
We derive a peak 2~cm flux density of 1.6~mJy/beam for this galaxy.
\citet{stawar86} have imaged this object at 4{\arcsec}
resolution at 6~cm and 20~cm.  Both maps show extended jet-like 
structure in P.A. $\sim$~45{\arcdeg}, more or less along
the minor axis of the host galaxy. They found a peak 20~cm flux density of
11.4~mJy/beam, and a steep spectrum ($\alpha$ = $-$0.6) over the jets and 
at the core. 
The steep-spectrum jets may dominate the core emission within 
the central 4{\arcsec} beam, so the core could potentially have a flat 
spectrum at higher resolution. In the absence of higher resolution
maps, we list this object as a steep spectrum source in Table 1.

\paragraph{NGC~5005}
\citet{vilet90} find S$_{20cm}^{peak}$ = 14.6~mJy/beam
and S$_{6cm}^{peak}$ = 2.7~mJy/beam at resolutions of $\sim$1{\farcs}1
and $\sim$1{\farcs}0, respectively (a spectral 
index of $\alpha^{20}_{6}$~= $-$1.8). The source is resolved at
both wavelengths, in P.A. $\sim$~135{\arcdeg} on a 
$\sim$5{\arcsec} scale, again more or less along
the minor axis of the host galaxy.
\citet{humet85} find a 20~cm peak flux-density of 70~mJy/beam
at 14{\arcsec} resolution, with extended emission
(about 2{\arcmin} extent) in P.A. 67{\arcdeg}, along
the major axis of the galaxy. FIRST also lists this source with a peak 20~cm 
flux density of 35.9~mJy/beam (at 5{\arcsec} resolution),
with extended emission in P.A. 124{\arcdeg}.
The relatively steep spectral slope between 20 and 6~cm,
and the extended radio morphology at these frequencies,
is consistent with our 2~cm non-detection.

\paragraph{NGC~5033}
FIRST lists a peak flux-density of 1.4~mJy/beam at 20~cm
at a resolution of 5{\arcsec}.
The inner radio emission in the FIRST map
is extended in P.A. 91{\arcdeg}, along the minor axis 
of the host galaxy, and there is an outer
diffuse component which is more aligned with the
host galaxy major axis.
Combining the FIRST peak flux-density with our 2~cm peak flux-density 
(1.4~mJy/beam) results in a non-simultaneous spectral index 
$\alpha^{20}_{2}~\geq$ 0. 

\paragraph{NGC~5055}
The 20~cm FIRST map has peak flux-density 1.5~mJy/beam,
and shows extended emission in P.A.  42{\arcdeg}.
We did not detect this galaxy at 2~cm.

\paragraph{NGC~5273}
FIRST lists a peak flux-density of 2.6~mJy/beam at 20~cm;
the extended radio emission (P.A. 180{\arcdeg}) is along
the major axis of the host galaxy disk. We did not
detect this galaxy at 2~cm.

\paragraph{NGC~5866}
\citet{conet90} find a peak 20~cm flux density of 9.5~mJy/beam in a
5{\arcsec} resolution map made with the VLA. 
This is somewhat different from the value listed in
FIRST - 20~cm peak flux-density of 15.4~mJy/beam at a 
resolution of 5{\arcsec}. \citet{wrohee91} derived
a flux density of 7.4~mJy from their 5{\arcsec} resolution
6~cm map. We measured a peak 2~cm flux density of 7.1~mJy/beam at 
0{\farcs}15 resolution, so the core has a spectral
index, $\alpha^{6}_{2}~\geq$ 0. The extended flux density in
the FIRST map is in P.A. 132{\arcdeg}, along
the major axis of the host galaxy,
not unexpected in this transition object.
This is the only transition object in the sample in which
we detected a 2~cm radio core.

\paragraph{NGC~6500}
The spectrum of the core of this spiral is known to be flat on 
scales of $\sim$20 mas \citep{jonet81}.
The 1{\farcs}2 resolution 20~cm map of \citet{unget89}
shows extended emission in P.A. 140{\arcdeg},
along the minor axis of the galaxy,
suggestive of an outflow. Their higher resolution
(0{\farcs}3) 20~cm MERLIN map 
shows an extension in P.A. $\sim$70{\arcdeg}.

\paragraph{NGC~7217}
Our simultaneous 3.6~cm observation on 1996 October 11,
with resolution 0{\farcs}27 x 0{\farcs}25, detected
an unresolved nuclear source with flux density 1.2~mJy. 
Since the 3.6~cm detection gives us an accurate position for the 
radio core, we use a 5$\sigma$ upper limit of 0.6~mJy at 2~cm.

\singlespace

\singlespace
\pagestyle{empty}
\clearpage

\begin{figure}
\vspace{-2in}
\figurenum{1}
\epsscale{1.0}
\plotone{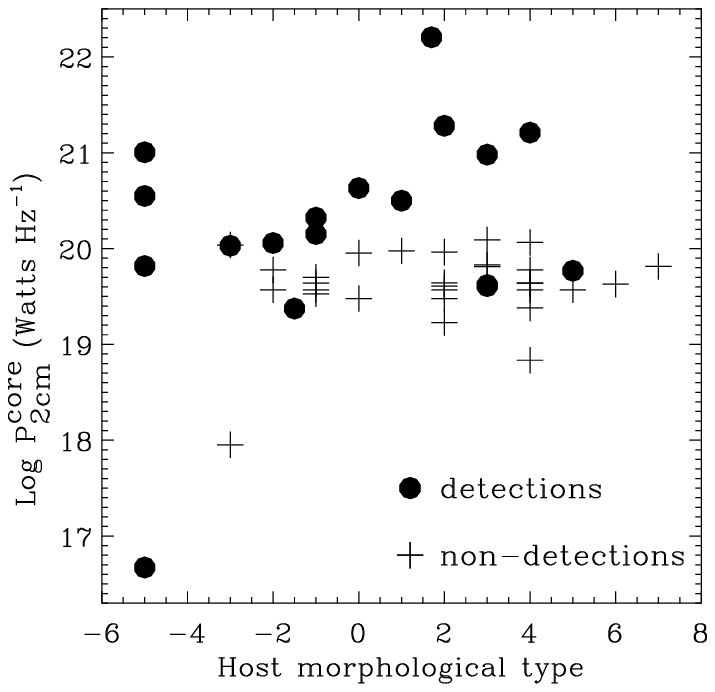}
\vspace{-4in}
\caption{
Distribution of the 2~cm nuclear power for all 48 LLAGNs in the sample, as
a function of host galaxy morphological type. Detections are 
plotted with filled circles while non-detections are plotted with
crosses. Note the large number of spiral galaxies detected.
}
\end{figure}
\clearpage

\begin{figure}
\figurenum{2}
\epsscale{1.0}
\plotone{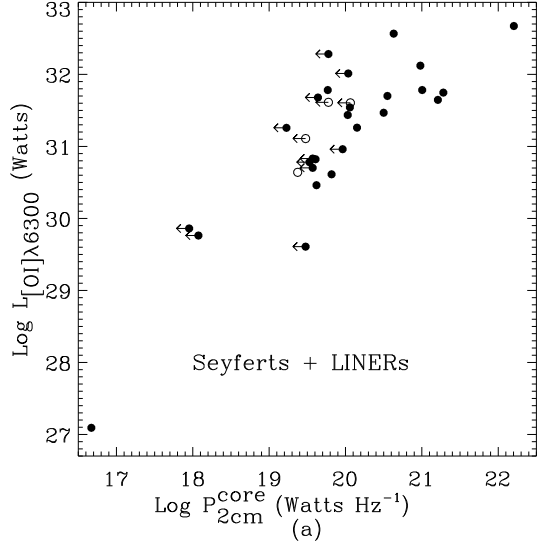}

\vspace{-6in}

\plotone{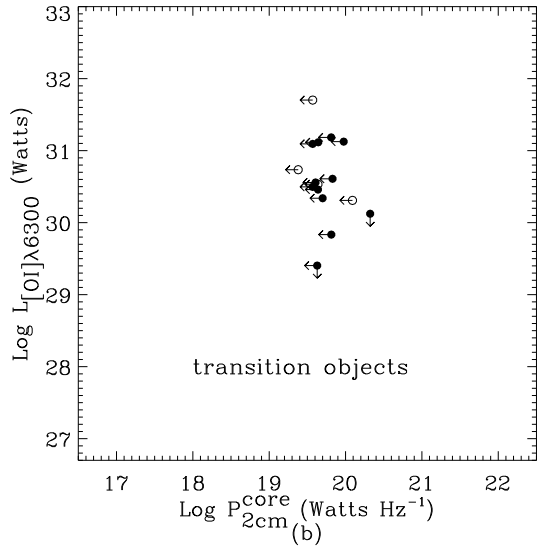}

\vspace{-5in}
\caption{
(a)~Relationship between [O~I]~$\lambda$6300{\AA} luminosity and 2~cm radio 
power for Seyfert and LINER nuclei in our sample. 
Objects with non-photometric [O~I]~$\lambda$6300{\AA} flux measurements 
are shown as open circles;
(b)~same as (a) but for transition nuclei in our sample.
}
\end{figure}

\begin{figure}
\figurenum{3}
\epsscale{1.0}
\plotone{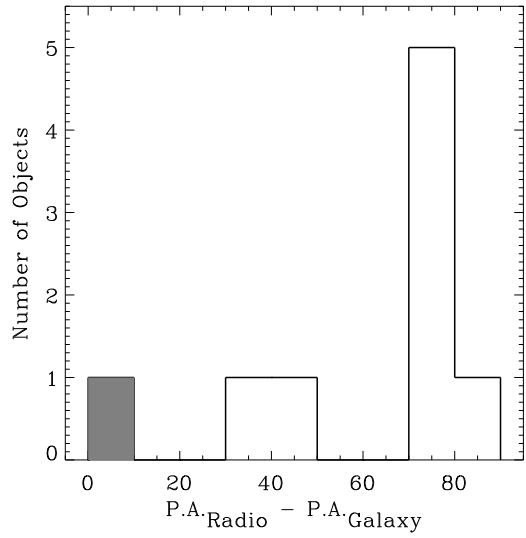}
\vspace{-3in}
\caption{
Histogram of the distribution of the difference
between the position angle of the 20~cm radio emission at 
1{\arcsec}--5{\arcsec} resolution and the position
angle of the host galaxy major axis, as listed in the UGC, for all
LLAGNs in our sample which contain a 2~cm compact core and have
extended nuclear radio emission.
The single transition object is shown in grey.
}
\end{figure}

\clearpage

\hoffset=-1.0in
\voffset=-2in
\begin{figure}
\epsscale{1.35}
\plotone{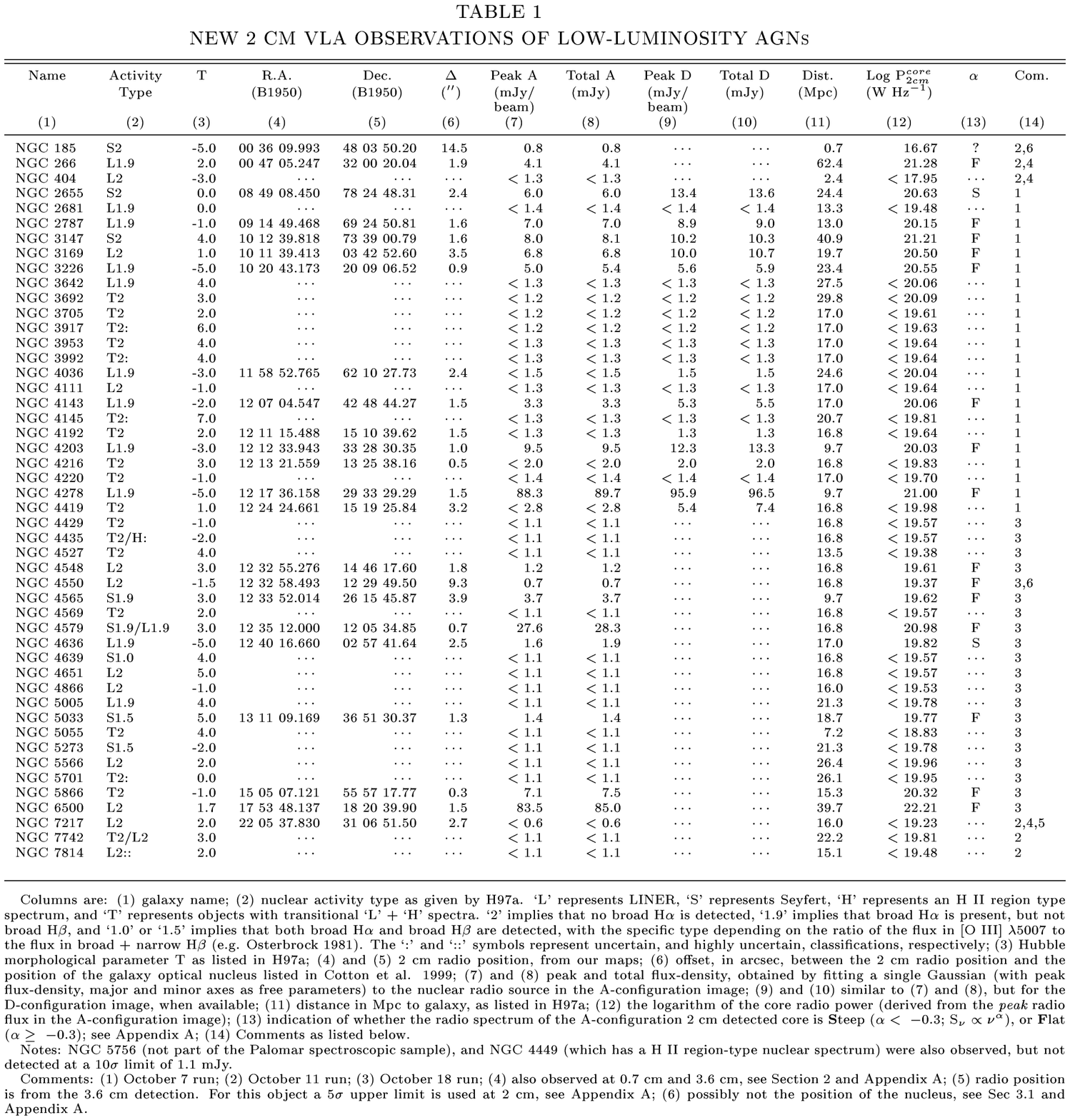}
\end{figure}

\clearpage

\begin{figure}
\epsscale{1.5}
\plotone{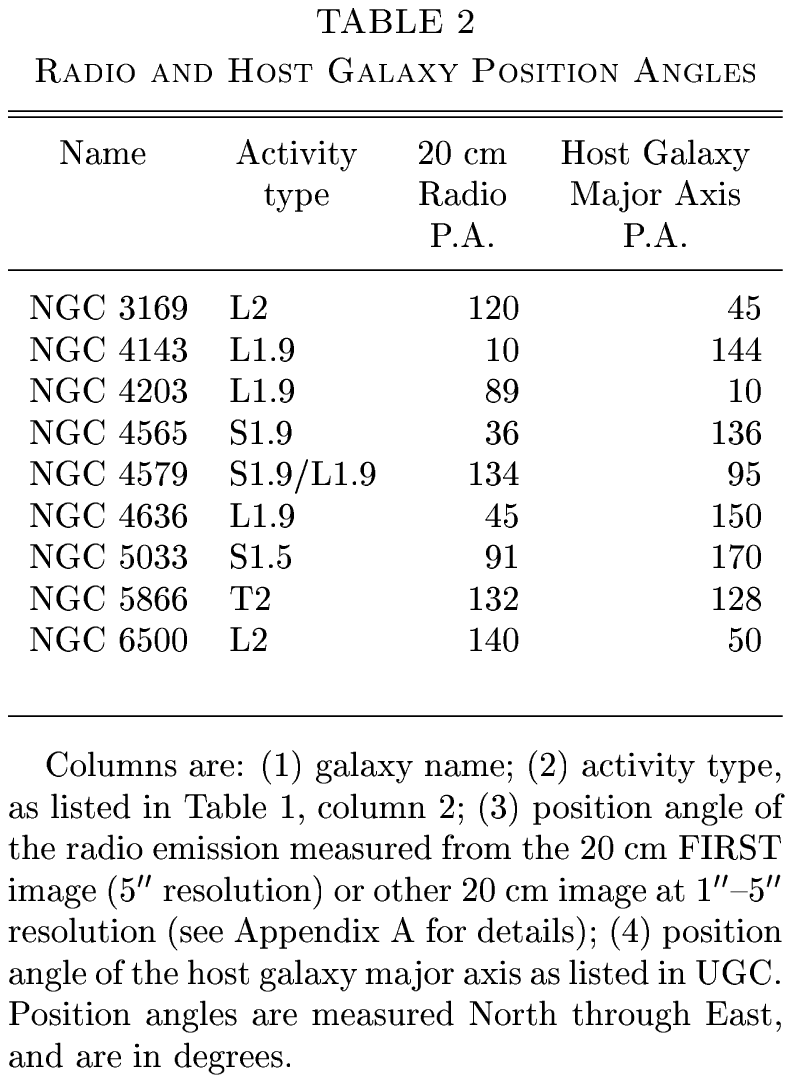}
\end{figure}

\end{document}